\documentclass{webofc}
\usepackage[varg]{txfonts}   
\usepackage{hyperref}
\usepackage{url}
\usepackage{graphicx}
\renewcommand{\vec}[1]{\mbox{\boldmath $#1$}}
\hypersetup{colorlinks=true,citecolor=blue,urlcolor=blue,linkcolor=blue}
\begin{document}
\title{Microscopic description of induced fission 
in a configuration interaction approach
}
%
%

\author{\firstname{Kotaro} \lastname{Uzawa}\inst{1}\fnsep\thanks{\email{uzawa.kotaro.37s@st.kyoto-u.ac.jp}} \and
        \firstname{Kouichi} \lastname{Hagino}\inst{1}
        \and
        \firstname{George} \lastname{F. Bertsch}\inst{2}
}

\institute{Department of Physics, Kyoto University, Kyoto 606-8502, Japan
\and
 Department of Physics and Institute for Nuclear Theory, University of Washington, Seattle, Washington 98915, USA
}

\abstract{%
Even though more than 80 years have passed since the discovery of fission, its microscopic understanding has still been unclear. 
To clarify the underlying mechanics of induced fission, we analyze the distribution of a fission width 
using a miscropic framework based on a configuration-interaction approach. 
The distribution is known to follow a chi-squared distribution, which is characterized by 
the effective number of decay channels, $\nu$. 
We introduce an effective Hamitonian for the space of compound nucleus states and 
estimate $\nu$ from 
the rank of the imaginary part of the effective Hamiltonian.   
Applying the model to $^{235}$U(n,f), 
we succesfully reproduce the empirical value of $\nu=2.3\pm1.1$. 
We also find that 
$\nu$ is insensitve to the number of fission channels, which is consistent with an experimental finding.
}
\maketitle
\section{Introduction}
\label{intro}
Nuclear fission was discovered in 1938\cite{Hahn}, 
and in the following year, 1939, Bohr and Wheeler explained its mechanism based on 
the liquid-drop model\cite{Bohr}. That is, fission takes place by overcoming a fission barrier, 
which is formed as a consequence of the competition between the surface and the Coulomb energies. 
The basic idea of such macroscopic picture has still been valid until now, 
and many dynamical fission models 
assume a nuclear shape evolution in a potential energy surface. 
On the other hand, a microscopic understanding of nuclear fission has still been 
unclear \cite{bender2020}.
In a fission process, 
collective and single-particle degrees of freedom interact with each other in a complex way, 
which makes a microscopical treatment of a fission process formidably 
difficult.
In recent years, with the help of developments of the nuclear many-body theory and computing powers, 
there have been many attempts to microscopically understand nuclear fission \cite{bender2020}. 

One of the most important 
quantities in nuclear fission is 
the distribution of a fission width. 
The distribution of decay widths of a compound nucleus is known to follow a chi-squared distribution, 
which is caracterized by the degrees of freedom, $\nu$\cite{Porter}.
The parameter $\nu$ contains information on a decay channel, and, if the output channels are independent of each other, 
$\nu$ is equivalent to the number of open channels\cite{Porter}.
In the case of fission of actinide nuclei, $\nu$ was found to be $O(1)$ \cite{Porter,Leal}. 
Notice that this value is considerably smaller than the number of output channels, 
which are characterized by quantum numbers and excitation energies of fission fragments.
Porter and Thomas explained this discrepancy based on the transition state theory \cite{Bohr}, for which $\nu$ 
was proven to be identical to the number of transition states.

In this paper, we shall analyze a smallness of $\nu$ based on a microscopic fission model.
Our model is based on the configuration-interaction approach with single-particle levels constructed with 
the constrained density functional theory with several nuclear shapes \cite{Bertsch1, Uzawa1}. 
We shall analyze a fission decay process of a compound nucleus state by 
constructing an effective Hamiltonian from our model Hamiltonian. 
This method provides a distribution of decay widths itself, and this is suitable 
for analyzing the degrees of freedom in a fission channel.
We mention that this is the first microscopical analysis of a distribution of a fission width. 
While the value of $\nu$ has been understood phenomenologically by now based on the transition state theory, 
the present work will offer insights into how a fission process takes place 
at a microscopic level.

\section{Method}

To describe coherently a collective deformation and single-particle excitations during a fission, 
we follow the idea of the generator-coordinate method(GCM) \cite{Ring}, in which a many-body wave function is constructed 
as,
\begin{equation}
|\Psi\rangle=\int dQ\sum_{\mu}\, f(Q,E_\mu)|Q,E_\mu\rangle. 
\label{GCM}
\end{equation}
Here $|Q,E_\mu\rangle$ denotes a mean-field wave function labeled by the deformation parameter $Q$ 
and the particle-hole excitation energy $E_\mu$.
Using such GCM basis, the Hamiltonian matrix is represented as $H_{k\mu,k'\mu'}=\langle Q_k,E_\mu|H|Q_{k'},E_{\mu'}\rangle$.
For residual interactions, we employ a monopole pairing interaction,  
\begin{equation}
  H_{\rm pair}=-G\sum_{i\neq j}a^{\dagger}_{i}a^{\dagger}_{\bar{i}}a_{\bar{j}}a_{j},   
\end{equation}
as well as  a diabatic interaction \cite{Hagino1},
\begin{eqnarray}
\frac{\langle Q,E_\mu|v_{db}|Q',E_{\mu'}\rangle}{\langle Q,E_\mu|Q',E_{\mu'}\rangle}
=\frac{E(Q,E_\mu)+E(Q',E_{\mu'})}{2} \nonumber +h_2{\rm ln}(\langle Q,E_\mu|Q',E_{\mu'}\rangle),  
\end{eqnarray}
where $G$ and $h_2$ are the interaction strength parameters.
The Hamiltonian matrix has diagonal sub-blocks characterized by $Q$, which we call $Q$-blocks.
The left-most and the right-most $Q$-blocks correspond to the compound states and the pre-fission configurations, 
respectively.
Both of them are at high excitation energies, and we thus replace them by random matrices based on a Gaussian orthogonal 
ensemble (GOE), which is characterized by an interaction strength $v$ and the matrix dimension, $N_{\rm GOE}$. 
The right-end scission configurations have couplings to the continuum states with two fission fragments,  
and an imaginary matrix $-i\Gamma_{\rm fis}/2$ is added to the Hamiltonian matrix \footnote{
A similar imaginary magrix, $-i\Gamma_{\rm cap}/2$, may be added also to the left-end GOE matrix \cite{Bertsch1, Uzawa1}, 
but this is not relevant to the discussion on the effective Hamiltonian (see below). }. 
Then, the Hamiltonian matrix has the following structure,
\begin{equation}
\label{3D-H0}
H  = \left(\begin{matrix}
     H^{(L)}_{\rm GOE} & (V^{(L)})^T &  & & & \cr
     V^{(L)} & H_1 & V_{1,2} &  & \text{\large{\textit{O}}}& \cr
      & V_{2,1} & H_2 & V_{2,3} &  & \cr
      &  &  &\ddots &  & \cr
      \text{\large{\textit{O}}}&  & & V_{N-1,N}& H_{N} & (V^{(R)})^T \cr
      &  & & & V^{(R)} &\tilde{H}^{(R)}_{\rm GOE} \cr
           \end{matrix}\right), 
\end{equation}
Here $O$ denotes the zero matrix, and 
$H^{(L)}_{\rm GOE}$ and $\tilde{H}^{(R)}_{\rm GOE}$ are the GOE matrices, the latter having the imaginary part.
The overlap matrix $N_{k\mu,k'\mu'}=\langle Q_k,E_\mu|Q_{k'},E_{\mu'}\rangle$ has a similar block structure.

Let us write this Hamitonian as, 
\begin{equation}
H  = \left(\begin{matrix}
     H^{(L)}_{\rm GOE} & (\vec{V}^{(L)})^T \cr
     \vec{V}^{(L)} & H_Q 
           \end{matrix}\right),
\end{equation}
with $(\vec{V}^{(L)})^T= \left( (V^{(L)})^T,O,O,\cdots O\right)$.  
The compound states in $H^{(L)}_{\rm GOE}$ decay via fission through 
the coupling to $H_Q$. We incorporate this effect by constructing an effective Hamiltonian as,
\begin{align}
H_{\rm eff}(E)&=
H^{(L)}_{\rm GOE}-\left(\vec{V}^{(L)}\right)^T(H_Q-EN_Q)^{-1}\vec{V}^{(L)} \notag\\
&\equiv H^{(L)}_{\rm GOE}+\Delta(E)-i\Gamma_{\rm eff}(E)/2,
\label{eHamiltonian}
\end{align}
where $\Delta(E)$ and $\Gamma_{\rm eff}(E)$ are the real and the imaginary parts of the self-energy, respectively. 
When $\Delta(E)$ can be treated perturbatively, 
the real part of $H_{\rm eff}$ approximately follows GOE, and thus 
the distribution of the decay width, that is, the imaginary part of the eigenvalues of $H_{\rm eff}$, 
follows approximately  the chi-squared distribution \cite{RMP}.
In this approximation, one could approximately regard 
the rank of $\Gamma_{\rm eff}$ as 
the degrees of freedom $\nu$. 
In this way, the value of $\nu$ in the fission channel can be microscopically estimated.

\section{Results}

Let us now apply the formalism to 
the induced fission of $^{236}$U, for which the empirical value of $\nu$ is $2.3\pm1.1$ \cite{Porter}.
For the DFT calculations, we employ the Skyrme UNEDF1 functional \cite{unedf}.  
We solve the Kohn–Sham equations in the cylindrical coordinate with the {\tt Skyax} code\cite{skyax}.
The pairing interaction is not taken into account in constructing the GCM basis $|Q,E_\mu\rangle$, 
but it is taken into account as a residual interaction in constructing the Hamiltonian 
matrix.
The strength is set to be $G=0.16$ MeV to reproduce the excitation energy of the 
first excited $0^+$ state of $^{236}$U.
On the other hand, the strength for the diabatic interaction is set to be 
$h_2=1.5$ MeV \cite{Bertsch1}.

\begin{figure}[htbp]
\centering
\includegraphics[width=8cm,clip]{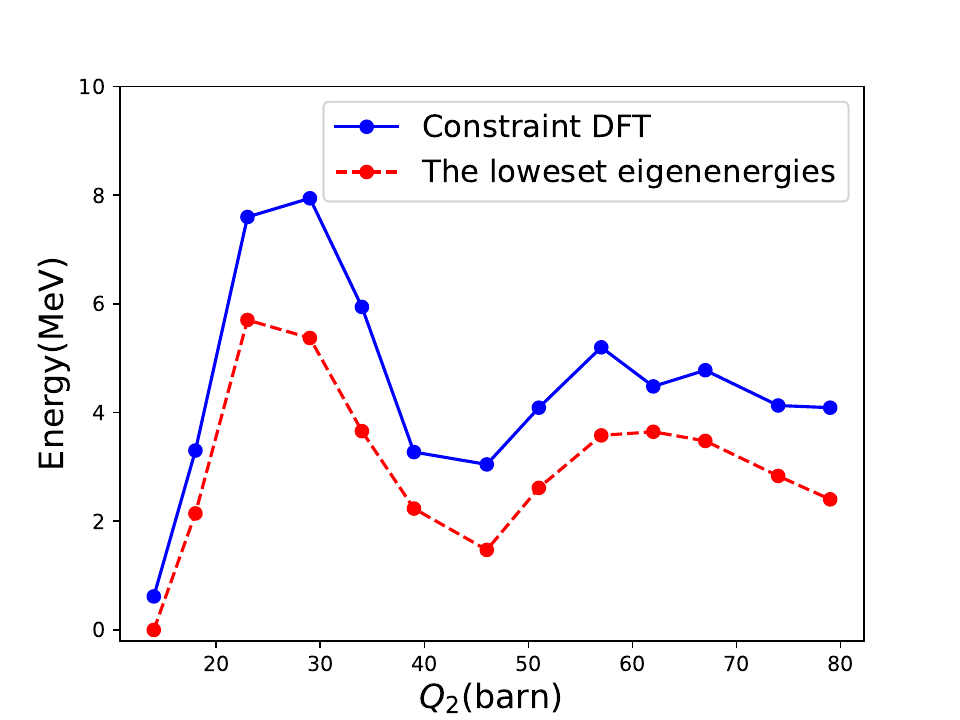}
\caption{
The fission barrier of $^{236}$U as a function of the mass quadrupole moment, $Q_2$.
The blue solid line shows the ground state energy at each $Q_2$. 
This is obtained with 
the DFT with the Skyrme UNEDF1 functional, after introducing the scaling factor of 0.71.  
The red dashed line shows the lowest eigenenergies obtained by diagonalizing the Hamiltonian at each $Q_2$.
}
\label{fig-1}      
\end{figure}

The fission barrier is calculated as a function of the mass 
quadrupole moment, $Q_{20}\equiv Q_2$, see the blue line in Fig. \ref{fig-1}. 
The figure also shows by the red dashed line 
the lowest eigen-energy at each $Q_2$ 
after diagonalizing the Hamiltonian for each $Q$-block.
Due to the lack of triaxial deformation, the first fission barrier is somewhat overestimated. 
We thus rescale the fission barrier by 0.71 so that the energy difference between the lowest 
eigen-energy at $Q_2=14$ b and that at $Q_2=23$ b becomes identical to the experimentally
determined barrier height, 5.7 MeV.

Based on the single-particle levels at each $Q_2$, we generate 
many-particle many-hole excited configurations at each $Q_2$. 
To this end, both proton and neutron excitations are taken into account up to 5 MeV.
As we have mentioned, we replace the left-end the right-end configurations at $Q_2=14$ b and 83 b, respectively, 
by GOE matrices. 
We take $N_{\rm GOE}=1000$ and $v=0.31$ MeV for the GOE matrices, 
which yield $\rho = v\pi/N^{1/2}_{\rm GOE}=31.8$ MeV$^{-1}$.  
We assume a diagonal width matrix, $\Gamma_{\rm fis}=\gamma_{\rm fis}\vec{1}$, where $\vec{1}$
is the unit matrix with the dimension of $N_{\rm GOE}$. 
We arbitrarily take $\gamma_{\rm fis}=0.015$ MeV based on the previous work \cite{Bertsch1}. 
Notice that the fission to capture branching ratio has been found insensitive to the size of the fission width\cite{Bertsch1, Uzawa2}, and the actual value of $\gamma_{\rm fis}$ would not be important. 

\begin{figure}[htbp]
\centering
\includegraphics[width=8cm,clip]{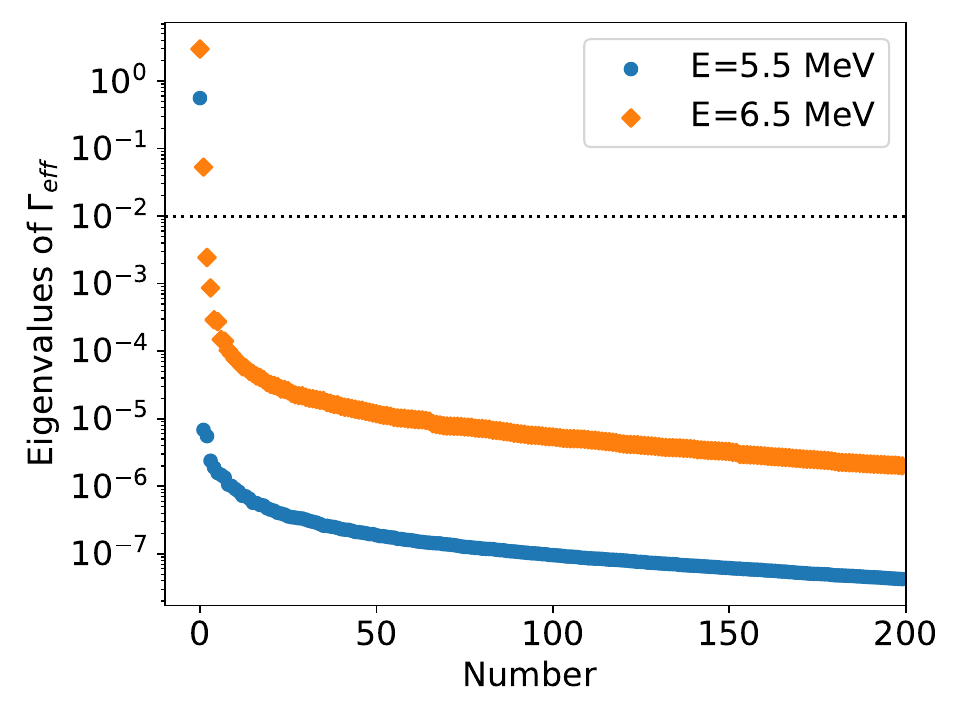}
\caption{Eigenvalues of $\Gamma_{\rm eff}$ at $E$=5.5 MeV (the blue circles) 
and 6.5 MeV (the orange diamonds) for a typical sample. 
They are plotted in the descending order.
The dotted line shows the threshold value $10^{-2}$ to be used in calculating the rank of $\Gamma_{\rm eff}$.
}
\label{fig-2}      
\end{figure}

After we calculate the GCM kernels, $H$ and $N$, the effective Hamiltonian is constructed according 
to Eq.(\ref{eHamiltonian}). 
In order to numerically determine the rank of $\Gamma_{\rm eff}$,
one needs to set a threshold value 
so that the rank is defined as the number of eigenvalues which are larger than it. 
To determine the threshold value, we analyze the distribution of the eigenvalues of $\Gamma_{\rm eff}$ 
at $E$=5.5 and 6.5 MeV (see Fig.\ref{fig-2}).
One can see that there is a clear gap, especially at $E=5.5$ MeV, between 
large eigenvalues and negligibly small eigenvalues. 
In the following calculations, we shall set $10^{-2}$ for the threshold value, as  is 
indicated by the dotted line in Fig.\ref{fig-2}. 

\begin{figure}[htbp]
\centering
\includegraphics[width=8cm,clip]{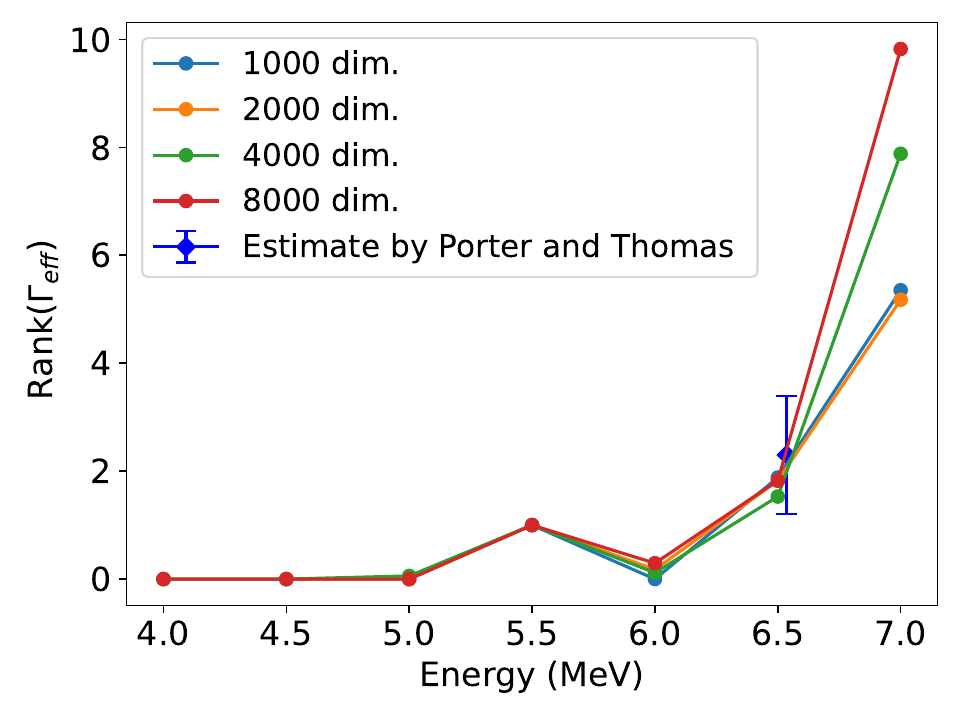}
\caption{
The rank of $\Gamma_{\rm eff}$ as a function of excitation energy, obtained after averaging over 100 samples. 
The different lines indicate the results of 
different dimensions of the right-end GOE matrix. 
The blue diamond shows the empirical estimate of $\nu$ \cite{Porter}. 
}
\label{fig-3}      
\end{figure}

The energy dependence of the rank so determined is shown in Fig. \ref{fig-3}.
The blue line shows the result with $N_{\rm GOE}=1000$ as a reference.
It is remarkable that 
our calculation reproduces a small number of $\nu={\rm rank}(\Gamma_{\rm eff})$, 
the value of which is consistent with the empirical value, $\nu=2.3\pm1.1$, at $E=6.536$ MeV \cite{Porter}.
Notice that ${\rm rank}(\Gamma_{\rm fis})$ tends to increase as the excitation energy increases. This 
reflects the fact that channels gradually open as the excitation energy increases.
We also find that at low energies 
${\rm rank}(\Gamma_{\rm eff})$ is insensitive to the size of the right-end GOE matrix, that is, $N_{\rm GOE}$, 
as shown in Fig. \ref{fig-3}.
This is compatible with the experimental finding that the number of apparent fission channels 
is not related to the degrees of freedom $\nu$.

\section{Summary}

Using the microscopical fission model based on the DFT and the CI approach, we have discussed the distribution 
of a fission decay width of compound nucleus states of $^{236}$U. 
In our model, many-body configurations are generated according to the mass quadrupole moment $Q_2$. 
To analyze the fission decay width of compound nucleus states, we have 
constructed an effective Hamiltonian which acts on the space of the compound nucleus states, and 
identified its imaginary part, $\Gamma_{\rm eff}$, as the fission decay width. 
We have found that $\Gamma_{\rm eff}$ has a small number of non-negligible eigenvalues, 
and thus the rank of $\Gamma_{\rm eff}$ is small. 
As long as the real part of the self-energy is small, 
the rank of $\Gamma_{\rm eff}$ is identical to the effective number of degrees of freedom, $\nu$. 
We have shown that the estimated value of $\nu$ is consistent with the empirical value, $\nu=2.3\pm1.1$.
Furthermore, we have also shown that the rank of $\Gamma_{\rm fis}$ is insensitive to the dimension of the right-end GOE 
matrix. 
This is consistent with the experimental finding that the apparent number of the exit channels is not reflected 
in the number of degrees of freedom.

We emphasize that this is the first microscopic estimate of the number of degrees of freedom in a fission channel. 
A more detailed discussion on the correspondence between $\nu$ and many-body wavefunctions will be given in 
a separate publication. 

\section*{Acknowledgement}
This work was supported in part by JSPS KAKENHI
Grants No. JP19K03861, JP23K03414, and JP23KJ1212.

\end{document}